# Message-Passing Multi-Cell Molecular Dynamics on the Connection Machine 5

D. M. Beazley* and P. S. Lomdahl†

{dmb,pxl}@viking.lanl.gov

*Theoretical Division and Advanced Computing Laboratory*
*Los Alamos National Laboratory, Los Alamos, New Mexico 87545*

**ABSTRACT**

We present a new scalable algorithm for short-range molecular dynamics simulations on distributed memory MIMD multicomputer based on a message-passing multi-cell approach. We have implemented the algorithm on the Connection Machine 5 (CM-5) and demonstrate that meso-scale molecular dynamics with more than $10^8$ particles is now possible on massively parallel MIMD computers. Typical runs show single particle update-times of $0.15\mu s$ in 2 dimensions (2D) and approximately $1\mu s$ in 3 dimensions (3D) on a 1024 node CM-5 without vector units, corresponding to more than 1.8 GFlops overall performance. We also present a scaling equation which agrees well with actually observed timings.

# 1 Introduction

The molecular dynamics (MD) method [1, 2, 3] has been known for several decades and has been used successfully in atomistic simulation models of thousands of interacting particles to describe structural and dynamical properties of simple physical systems, such liquids and solids. However, the application of MD to realistic problems from materials science and condensed matter physics has until recently been hampered by the availability of sufficiently powerful computers. Simulations with only a few thousand atoms are generally too small to capture realistic densities of defects like dislocations and microstructural grain boundaries. Similarly, it is often necessary to follow the time evolution of of an MD simulation for hundred of *ns* if laboratory-scale strain-rates are to be realistically modeled. At normal densities in solids a cube of $10^3 \times 10^3 \times 10^3 = 10^9$ atoms of material is roughly $0.35 \mu m$ on a side. Even though this may seem like a very small piece of material, simulating its dynamical properties for even a few *ns* presents a formidable problem for conventional MD simulations today. However, new generations of supercomputers are rapidly changing this picture.

Simulations with $10^6$ atoms have been performed on conventional vector supercomputers [5]. However, it is parallel multicomputers that seem to hold the greatest potential for reaching beyond $10^9$ atoms. Coarse-grained transputer systems have been used in successful 2D simulations with $10^6$ atoms [6]. Data parallel SIMD implementations on the CM-2 have also recently been presented [7, 8] which have the potential for multi-million particle simulations. In fact, recent work by Holian and collaborators [9, 10] have demonstrated 2D simulations for $1.6 \times 10^7$ atoms on a $64K$ processor CM-200. Work on MIMD implementaions of large scale MD is now also starting to appear [11, 12].

In this paper we present a scalable parallel MD algorithm which allows for the simulation of at least $10^8$ particles interacting via a relative short range potential. We have implemented this algorithm on the Connection Machine 5 (CM-5) from Thinking Machines Corporation (TMC) and thus demonstrate that near macroscopic scale MD simulations are possible on today's state-of-the-art parallel multicomputers.

The paper is organized as follows: In section 2 we review the basic principles behind the physics of MD simulations, in section 3 we briefly outline the CM-5 architecture and its message-passing communication model relevant for our MD implementation. In section 4 we describe our MD algorithm and in section 5 we present our timing results with a scaling equation that can be used to predict the times. Finally, in Section 6 we conclude by looking at prospects for the future.

# 2 Short-range molecular dynamics

The classical MD method concerns the solution of Newton's equation of motion for $N$ interacting particles. This general $N$-body problem involves the calculation of $N(N-1)/2$ pair interactions in order to compute the total force on any given particle:

$$m_i \frac{d^2 \mathbf{r}_i}{dt^2} = - \sum_{j \neq i} \frac{\partial V_{ij}(|\mathbf{r}_j - \mathbf{r}_i|)}{\partial \mathbf{r}_i}, \quad (1)$$

here $\mathbf{r}_i$ indicates the instantaneous position and $m_i$ the mass of particle $i$. The complexity of the force calculation can be simplified considerably if the potential $V_{ij}(r)$ has a finite range of interaction. This is a reasonable approximation of the atomistic interactions in many solids and fluids. In this paper we will assume that this is the case and will be concerned exclusively with this short-range approximation. In the case where long-range interactions (like electrostatic and gravitational forces) can *not* be neglected, efficient parallel algorithms based on hierarchical tree structures are also becoming available [13].

As the prototypical short-range interaction potential we have taken the Lennard-Jones spline (LJ-spl) potential [9]. This potential has the usual Lennard-Jones 6-12 form modified with a cubic spline that makes the potential go exactly to zero at a distance $r_{max}$.

$$V(r) = \begin{cases} 4\epsilon \left( \left(\frac{\sigma}{r}\right)^{12} - \left(\frac{\sigma}{r}\right)^6 \right) & 0 < r \leq r_{spl} \\ -a_2 (r_{max}^2 - r^2)^2 + a_3 (r_{max}^2 - r^2)^3 & r_{spl} < r \leq r_{max} \\ 0 & r_{max} < r \end{cases} \quad (2)$$

Here $\sigma$ and $\epsilon$ are the usual Lennard-Jones parameters, $a_2$, $a_3$ and $r_{max}$ are given by the following expressions

$$r_{max}^2 = r_{spl}^2 \left\{ 5 - 5 \left[ 1 - \frac{1}{25} \left( 9 - \frac{24V}{r_{spl}V'} \right) \right]^{\frac{1}{2}} \right\}, \tag{3}$$

$$a_2 = \frac{5r_{spl}^2 - r_{max}^2}{8r_{spl}^3(r_{max}^2 - r_{spl}^2)} V' \ , \ a_3 = \frac{3r_{spl}^2 - r_{max}^2}{12r_{spl}^3(r_{max}^2 - r_{spl}^2)^2} V', \tag{4}$$

which are obtained by requiring that the force and potential are continuous and the derivative of the force vanish at the inflection point, $r_{spl} = 1.244455\sigma$. The cut-off distance is $r_{max} = 1.711238\sigma$. Unlike truncated LJ-potentials [7], the LJ-spl potential has a continuous derivative at $r_{max}$.

The number of interacting neighbors for each particle depends on the value of the cut-off distance $r_{max}$ and the particle density $\rho$. In 2D, the number of interacting neighbors is $N_{2D} = \pi r_{max}^2 \rho$ and in 3D, $N_{3D} = \frac{4}{3}\pi r_{max}^3 \rho$. For the LJ-spl potential above, this corresponds to approximately $9\sigma^2\rho$ neighbors in 2D and $21\sigma^3\rho$ in 3D. We should note that our algorithm is not dependent on the exact nature of the potential. More sophisticated many-body potentials like EAM [9] can be accommodated too.

## 3  Message-passing and the CM-5

The CM-5 [14] is a massively parallel multicomputer containing between 32 and 16384 processing nodes (PN), each of which consists of a 33 MHz SPARC processor, up to 32 Mbytes of memory and an optional 128 MFlop vector-processing unit (VU) for 64 bit floating-point and integer arithmetic. The PNs are controlled from one of several control processors (CP), each of which is a SUN Microsystems workstation. The PNs and the CP are interconnected via three high-bandwidth, low-latency networks: a control network, a data network, and a diagnostic network. The control network can perform scans (parallel prefix and suffix operations), reductions (max, add, OR, etc.), broadcasting and synchronization functions. The data network provides simultaneous point-to-point communication between multiple PNs at high bandwidth (over 5 Mbytes/sec/node). The network interface is memory mapped into user address space, so that operating system calls are avoided when it is accessed. The diagnostics network which allows for low-level access to system hardware (such as I/O or tests of system integrity) is privileged and access must be done via system calls.

The CM-5 is not a SIMD machine as its predecessor the CM-2. The programming model promoted by TMC is SPMD "single program - multiple data streams" or *synchronized* MIMD and is designed for data-parallel programming. However, it also supports very well the message-passing model that we adopt here for our MD algorithm. The CP broadcasts a single program over the control network to the PNs where it is executed locally. Because a program is usually much shorter than the instruction stream it generates, it is advantageous to just broadcast the program to the PNs rather than sending the entire instruction stream. This frees up additional network bandwidth for communicating user data. Most MIMD multicomputers communicate among themselves using either message-passing techniques or shared memory, but often there is little or no architectural support for synchronizing and coordinating sets of processors. These tasks are often left to the user who must implement elaborate protocols in order to assure processor synchrony. The CM-5 however, provides fast barrier synchronization in hardware via the control network. This is essential in the MD algorithm, where the time integration of the equations of motion can not proceed until the total force acting on each particle has been accumulated. Similarly, when PNs communicate via message-passing, program correctness demands that the processors know when it is safe to proceed after the communication step. For example, a given PN may not know whether it is done receiving messages or if more messages are pending from other PNs. In order for the PNs to know when message routing has terminated on the data network the CM-5 provides a fast "router-done" synchronization mechanism.

Access to the communication and data network is provided via the message-passing library CMMD. The library allows for both synchronous send-and-receive type communication as well as asynchronous (nonblocking) message-passing. We make extensive use of both types of communication models in our MD algorithm. Our code is written in ANSI C with the CMMD calls isolated in a few functions [1]. Adopting the code to other MIMD architectures supporting asynchronous messages passing should be a relatively straightforward process.

While the high level data-parallel programming model supported by C* and CMF certainly is useful for many applications, we have found that the larger degree of freedom provided with the direct message-passing model maps very cleanly on to our MD algorithm. There are obvious steps in our MD algorithm which have direct analogies in the hardware of the control and data networks of the CM-5, thus making use of CMMD very natural. We feel that the greater flexibility provided by the detailed control of our data layout and communications is essential for attaining high performance. While compiler writers for data-parallel languages do a commendable job in masking the complexity of the hardware for the user, it is our experience that the ultimate performance is usually achieved by a careful analysis of the physics and its mapping to a programming model closer to the hardware.

The 1024 node CM-5 installed at the Advanced Computing Laboratory (ACL) at LANL did not have VUs at the time of this writing and each SPARC PN only had 16 Mbytes of memory. In section 6 we will return to estimates of performance enhancements expected when the ACL's CM-5 is fully configured.

## 4 The multi-cell MD algorithm

We illustrate our algorithm in 2D. The method naturally extends to 3D.

### 4.1 Data structures

We consider space to be a rectangular region with periodic boundary conditions. This region is then subdivided into large cells which are assigned to the PNs on the CM-5. Particles are assigned to processors geometrically according to the particle's coordinates. For a 16 processor CM-5, space would be subdivided as shown with solid lines in Fig. 1.

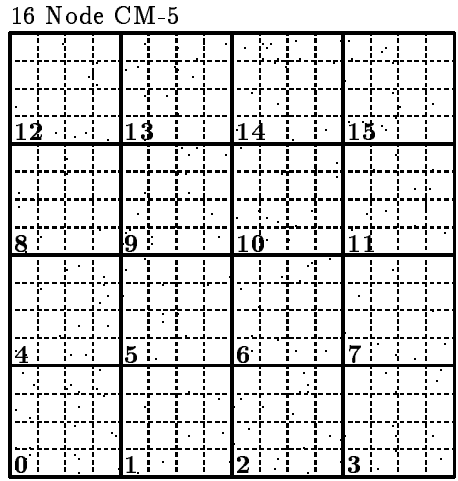

Figure 1. Processor and cell assignment.

The numbers in each region indicate the PN that would be assigned to that region of space. Points correspond to a sample set of particles. For solids, the particles are usually uniformly distributed and each processor could be assigned hundreds to tens of thousands of particles. Due to memory constraints on the current CM-5 (16 Mbytes/PN), no more than about 100,000 particles can be "safely" assigned to any one PN.

---

[1] We have used the GNU C compiler, gcc, (Vers. 2.2.2) and the CMMD 1.0 library for all timings reported here. CMMD 2.0 was just becoming available at the time of this writing.

For a large set of particles, the region assigned to each processor will have dimensions significantly larger than the interaction cut-off distance, $r_{max}$. In general, there will be a large number of particles on each processor that do not interact with each other. We seek a method to organize the particles so that a particle's neighbors can be quickly located for the force calculation and so the number of interactions calculated is minimized. To solve the problem, the region on each processor is subdivided into a large collection of small cells. Each cell created is chosen to be slightly larger than the interaction cut-off. Particles are then assigned to an appropriate cell geometrically as before. It is important to note that number of cells created depends only on the size of the region and the interaction cut-off, $r_{max}$, not on the number of the PNs available. For large simulations it is possible to subdivide the space on each processor into thousands of cells. In Fig. 1, the dashed lines indicate the smaller cells created on each PN. In this case, 16 cells per processor are being used. Each cell has dimensions larger than the cut-off distance. The same number of cells are created on every PN.

This cell structure forms the foundation of our algorithm. For storage, each particle consists of a C structure containing the position, velocity, force, mass, and a particle type. For memory management, associated with each cell is a small block of memory where the particles are stored sequentially in a list. This method of storage is important for the communication aspects of our interaction calculation since we communicate entire cells, not individual particles. The sequential nature of the data allows us to easily communicate the entire contents of a cell by simply sending a small block of memory. In the current implementation the memory block is of a fixed size, but a dynamic allocation scheme can be implemented with only a minor effort.

## 4.2 Interactions

With the cell structure now in place, interactions can be effectively calculated. The calculation of forces on a single particle only involves the particles in the same cell and neighboring cells. The calculation of forces for all of the particles in a cell is a two step process. First, all of the forces between particles in the same cell are calculated. Next, forces from particles in the neighboring cells are calculated by following an interaction path [7] that describes how we compute the interactions with neighboring cells. In 2D the path is as shown in Fig. 2.

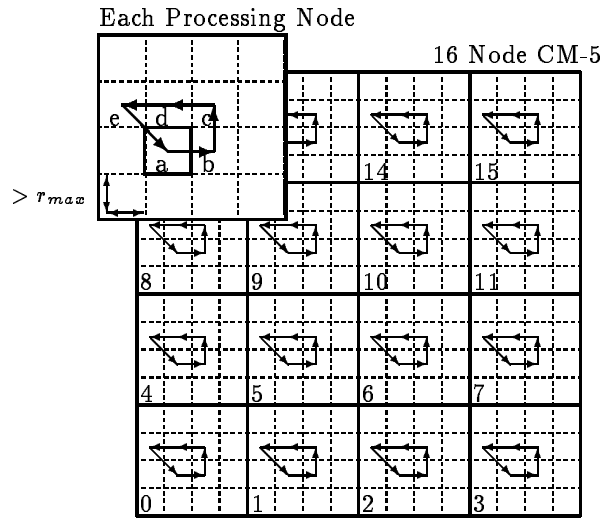

Figure 2. Force calculation.

Forces are being calculated for the particles in cell (a). Interactions between all particles in cell (a) are first calculated. Interactions with particles in neighboring cells are then calculated in the order shown (b-e). Once interactions with cell (e) are calculated, the process is complete. As we calculate forces with the neighboring cells the total force is accumulated by the original cell and the cells along the path (using Newton's third law). To calculate all forces, this procedure is carried out on all cells on all processors. Cell (a) will accumulate the interactions from its lower neighbors when they calculate their interactions.

On each PN, the calculation of forces proceeds sequentially through all of the cells. This process then occurs in parallel on all processors on the CM-5. Globally, the processors are operating in a loosely-synchronous mode where each processor is calculating the forces for the same internal cell at approximately the same time. For most cells on a processor, all of the neighboring cells are on the same processor. This allows most cells to calculate interactions without any communications. However, for cells along the edge of each processor, we must calculate interactions between cells on different PNs. When this occurs, the message-passing features of the CM-5 are utilized. Particles are sent to neighboring processors and received from other processors. When communication is necessary, all of the PNs synchronize and participate in a send-and-receive type communication. Since each processor has the identical cell structure and are operating on their cells in the same order, when one processor needs to calculate an interaction with particles on another processor, all processors will have to do this. Each processor sends the particle data for the cell to the appropriate processor and simultaneously receives particles from another processor. The force calculation then proceeds with each processor operating on the particles it received. The key idea is that whenever a processor boundary is crossed along the interaction path, all processors participate in a synchronous communications step and the force calculation continues. It should be noted that synchronization only occurs in message-passing. At all other times, the processors are running asynchronously.

### Interaction Algorithm (Executes on all PNs simultaneously)

```
Variables:
    Cells(Xcells,Ycells) : 2d array of cells on the processor.
    Path[PathLength]     : Array of increments describing interation path.
    CellBuffer           : Temporary storage for a single cell
    CurrentCell          : Pointer to the cell for which interactions are being calculated.
    PathCell             : Current cell on interaction path.  Can be on a neighboring processor.
    InteractCell         : Same as PathCell except with periodic boundary conditions.
    ReceiveCell          : Pointer to where incoming particles are received during communications.
    HomeProcessor        : The ID number of this PN.

ComputeAllInteractions()
    For i = 0 to Xcells
        For j = 0 to Ycells {
            ComputeInteractionsSameCell(Cells(i,j))
            CurrentCell  = Cells(i,j)
            PathCell     = (i,j)
            InteractCell = (i,j)
            For k = 1 to PathLength {
                PathCell = PathCell + Path[k]
                InteractCell = InteractCell + Path[k]
                If (Processor(InteractCell) != HomeProcessor) {
                            Destination = Processor(InteractCell)
(1)                         Source      = Processor((Xcells,Ycells) - InteractCell)
(2)                         If (Processor(PathCell) = HomeProcessor)
                                    Then ReceiveCell = Cells(i,j)
                                    Else ReceiveCell = CellBuffer
                            SendAndReceive(CurrentCell-->Destination, Source-->ReceiveCell)
(3)                         CurrentCell = ReceiveCell
(4)                         InteractCell = InteractCell mod (Xcells, Ycells)
                }
                ComputeInteractions(CurrentCell, InteractCell)
            }
        }
```

```
end(ComputeAllInteractions)

ComputeInteractionsSameCell(C)
    For I = 1 to (NumParticles(C) - 1)
        For J = I+1 to NumParticles(C)
            F = ComputeForce(Particle(I), Particle(J))
            Particle(I).Force = Particle(I).Force + F
            Particle(J).Force = Particle(J).Force - F
end(ComputeInteractionsSameCell)

ComputeInteractions(Cell1, Cell2)
    For I = 1 to (NumParticles(Cell1))
        For J = 1 to (NumParticles(Cell2))
            F = ComputeForce(Particle(Cell1,I), Particle(Cell2,J))
            Particle(Cell1,I).Force = Particle(Cell1,I).Force + F
            Particle(Cell2,J).Force = Particle(Cell2,J).Force - F
end(ComputeInteractions)
```

---

(1) The source processor corresponds to the processor opposite of the destination processor.
(2) If the path leaves the processor, particles will be received from a neighboring processor and are stored in CellBuffer. If the path reenters this processor, particles that were sent out earlier are being sent back and are stored in their original location Cell(i,j). Forces that were calculated with particles on nearby processors are returned with the particles at this time.
(3) This switches the processing to the proper set of particles after communications.
(4) This imposes periodic boundary conditions on the interaction path. When particles are received from a neighboring processor, this forces the interaction path to wrap around to the other side to work with the received particles. In one sense, this processor takes over the interaction path from the neighboring processor.

---

The main feature of this algorithm is that each PN must simultaneously manage its own cells and cells received from its neighbors. To do this, two separate interaction paths are utilized (these are described by the variables PathCell and InteractCell). One path (PathCell) is used to manage the particles that belong to this PN. It may leave this processor when PN boundaries are crossed, but always keeps track of where particles are located at any particular time in the algorithm. The second path (InteractCell) describes the actual computations that need to be performed by this PN. The path is identical to the first path except that we impose periodic processor boundary conditions. When the interaction path crosses a PN boundary, new particles will be received from a neighboring PN across the opposite boundary. With processor periodic boundary conditions, the second path will properly describe the required interactions with the received cells and cells on this PN.

It was noted earlier that the particles on each cell are stored sequentially in memory. This allows us to send all of the particles in a particular cell by simply sending a block of memory through the data network. This type of sending is faster than sending one particle at a time that would be required if the particles in each cell were scattered throughout memory. We send all of the particle data in a cell at each communication. This includes position, velocity, force, and mass. This is inefficient in the sense that we are sending data that is unnecessary in the force calculation. However, we found that the extra overhead associated with sending only the necessary data actually slowed down the communications process. Performance might be improved with a slight modification to the data structures.

Our algorithm can also be viewed as a compromise between sending individual particles and a minimal message-passing scheme. In this latter scheme one could imagine all outgoing particles in a given PN be buffered and only sent when all particles tagged for a given destination processor are ready. However, such a scheme would be significantly more costly in terms of local memory and would probably prohibit runs with $10^8$ particles.

## 4.3 Time integration

The numerical integration step is a minor part of the overall computation process compared to the effort involved in calculating forces. We have implemented the Verlet algorithm and a stochastic Runge-Kutta algorithm. The Verlet algorithm is a multi-step integrator requiring only one force evaluation per time-step [1, 2]. Although the algorithm is a multi-step integrator, it has the advantage of being self starting since only the positions and velocities are needed initially.

A second-order Runge-Kutta scheme that has been adapted to solve stochastic differential equations for Langevin dynamics [15] has also been implemented. The advantage of this algorithm is that it allows simulations within the canonical ensemble which has constant temperature. The disadvantage is that it requires two force evaluations per time-step. With no noise, this method reduces to the standard 2nd-order Runge-Kutta method for solving ODE's. Higher order (and more accurate) Runge-Kutta methods could also be used at the expense of more force evaluations and memory. Predictor corrector methods are not easily implemented since they require the storage of information from previous time-steps. This information would have to be carried along with each particle which will slow down communications and other computations. In both cases, higher order integrators require more memory which becomes an issue for large simulations with many million particles.

## 4.4 Redistribution

After each time step, particles in the system will be moved to new positions. Since our algorithm is geometrically based, we must update all of the internal data structures to account for positional changes of the particles. A special redistribution function is called after every time step that checks the coordinates of each particle. If a particle is in the wrong cell, it is moved to the proper cell. This leads to the problem of effectively redistributing the particles with a minimal amount of computation time.

When moving particles between cells, the particle lists on each cell will become fragmented as particles are deleted from various lists and added to others. When a particle is added to a list, it should be stored in spaces that have been deleted, but the search for deleted particles should not be a sequential search through the array. This would waste valuable processing time. To solve the problem, a linked list of deleted particles is constructed for each cell. When a particle is deleted from a cell, it's location is added to the list. When a particle is added to a cell, it can instantly find an empty position by looking at the first entry in the linked list. The construction of lists is this case is efficient and provides a way to manage memory.

There are two cases to consider. The first case involves a particle simply changing cells on the same PN. In this case, the particle is deleted from one cell and added to the new cell. This only involves memory copying. The second case is when a particle needs to move to a cell on a new processor. In this case, the particle is deleted from its cell and sent to the new processor using message-passing. The redistribution process does not lend itself to using synchronized message-passing since each processor does not know how many particles will leave, how many will be received, and from what processors particles will arrive. For this reason, all PNs are put into an asynchronous message sending/receiving mode. If a particle needs to be sent to another processor, the particle is thrown onto the data network and the processor continues to check other particles (this is a nonblocking send operation). Each processor polls the network to see if any incoming particles are waiting. If so, they are retrieved off the network and stored in the appropriate cell. The process of checking particles proceeds serially for all particles on each PN. A processor many send or receive particles through the network at any time and in general all processors are operating independently. The entire process is complete when all processors have checked all of their particles and all messages have been retrieved off the data network.

### Redistribution Algorithm

```
RedistributeParticles()
STEP 1:  for i = 0 to Xcells
```

```
              for j = 0 to Ycells {
                  for k = 1 to NumberOfParticle(Cells(i,j)) {
                      if (Particle(k) in wrong cell)
                          NewCell = ComputeCell(Particle(k))
                          if (NewCell on this processor)
(1)                           AddParticle(Particle(k)-->NewCell)
                          else
(2)                           SendAsynchronous(Particle(k)-->Processor(NewCell))
                          Delete(Particle(k))
(3)                       AddToLinkedList(k)
                  }
(4)           While(Network has incoming particles) Do {
                  if (ParticleReceived)
                      Receive(TempParticle)
                      NewCell = ComputeCell(TempParticle)
                      AddParticle(TempParticle-->NewCell)
              }
          }

STEP 2:   Signal that this processor is done.
(5)       While(AllProcessorsNotFinished) Do
              While(Network has incoming particles) Do {
                  if (ParticleReceived)
                      Receive(TempParticle)
                      NewCell = ComputeCell(TempParticle)
                      AddParticle(TempParticle-->NewCell)
              }
end(RedistributeParticles)
```

---

(1) When adding particle to a new cell, the linked list of deleted particles determines the location for the new particle.
(2) This is a nonblocking send.
(3) The position of each deleted particle is added to a linked list maintained by each cell.
(4) Network is checked after processing each cell. This corresponds to checking the network every 10-15 particles that are processed depending on the number of particles per cell.
(5) Upon completion, each processor must continue to check for incoming particles until all processors have finished.

---

Our experience indicates that the redistribution process is efficient and accounts for a small portion of the overall iteration time. In many cases, a particle will not leave it's cell after each time step and the number of particles changing cells is small compared with the total number of particles in the system (this is of course dependent on the density in the system). Inter-processor communications in this procedure accounts for an even smaller portion of the overall time since most particles will simply move to a cell on the same processor. These facts, allow us to update the data structures for a very large system of particles very quickly. The redistribution algorithm rebuilds all necessary data structures. There is no searching for neighbors or the construction of neighbor lists [11]. There is also an advantage in that this procedure is called after every time-step so we are assured that the data is in the proper layout before each force calculation.

As mentioned earlier, the redistribution procedure tends to fragment data. Deleted particles are generally stored together with active particles (and are even sent to other processors in the force calculation). For this reason, it is important for other procedures to ignore the deleted particles. As a run proceeds there may be a proliferation of deleted particles in the system. For example, it may occur that *all* of the particles on a particular cell or processor leave. In this case, this processor is left with a large collection of deleted particles which are never cleaned up and which actually get sent to other processors in the force calculation (although nothing is ever calculated). This does not seem to have an adverse effect on the overall performance although the iteration time does tend to increase slightly. We have not written any code to perform garbage collection, but this could certainly be implemented if desired. A garbage collection procedure would only need to be called every few thousand time-steps so this would not adversely affect the performance of the algorithm.

## 5 Scaling and timing properties

In this section we examine the scaling and timing properties of our algorithm.

### 5.1 A scaling model

Suppose, for the purpose of analysis, that the particles are uniformly distributed and that the region assigned to each PN is square. Further suppose that this square region is subdivided into a collection of square cells of equal area and define the following variables.

$N$ = Number of particles on this processor.
$c$ = Number of cells in each direction.

We first determine the approximate number of interactions. The number of particles in each cell is given by $N_c = N/c^2$. The number of interactions calculated by each cell is then given by $I_s = N_c(N_c - 1)/2$ and $I_n = 4N_c^2$, where $I_s$ is the number of interactions from particles in the same cell and $I_n$ is the number of interactions calculated from the 4 neighboring cells during the force calculation. The total interactions per cell is then given by $I_c = 9N_c^2/2 - N_c/2$. Summing over all of the cells on the processor and substituting for $N_c$ the total number of interactions calculated is given by

$$I_t = c^2 I_c = \frac{9N^2}{2c^2} - \frac{N}{2}. \qquad (5)$$

Next, we determine the amount of message-passing required by the force calculation. We examine the interaction path and count up the number of times a processor boundary is crossed. The right edge contributes $2(c-1)$ passes, the top $2(c-2)$, the left $2(c-1)$ and the two top corners contribute 7 passes. The total number of message passes is given by

$$M_t = 6c - 1. \qquad (6)$$

The amount of data being sent on each message-pass depends on the size of each cell. The number of bytes per cell is given by $D_c = N_c \beta = N\beta/c^2$ where $\beta$ is the number of bytes per particle. The total amount of data sent is

$$D_t = (6c - 1)D_c = \frac{(6c - 1)N}{c^2}\beta. \qquad (7)$$

The total interaction time can now be approximated. Define the following variables :

$\alpha$ = Interaction time (sec/interaction).
$\delta$ = Data transfer rate (bytes/sec).
$\gamma$ = message-passing overhead for each send (sec).
$\beta$ = bytes per particle.

The total interaction time is given by the following

$$\begin{aligned} T &= I_t \alpha + M_t \gamma + \frac{D_t}{\delta} & (8) \\ &= \left[\frac{9N^2}{2c^2} - \frac{N}{2}\right]\alpha + (6c - 1)\left[\frac{N\beta}{c^2\delta} + \gamma\right]. & (9) \end{aligned}$$

All the constants can be defined in terms of machine architecture specifications such as the Flop rate and network transfer rate. They can also be determined empirically.

The choice of the number of cells has a dramatic effect on the overall performance. Looking at equation (5) for the total number of interactions, we see that the number $I_t$ decreases inversely as the square of $c$. This alone suggests that we should have as many cells as possible to reduce the number of interactions calculated. The minimum number of interactions occurs when the cell size is exactly equal to the cut-off distance (If the cells are smaller than the cut-off the force calculation will be incorrect).

Increasing the number of cells by increasing $c$ causes the number of message-passes to increase linearly, but the amount of data sent decreases as $1/c$. Generally the transfer time is significantly longer than the overhead of setting up a message-pass so it is an advantage to increase the number of cells in this case as well. If the overhead for initiating a message send is extremely high, then it may be a disadvantage to increase the amount of message-passing. This does not seem to be the case on the CM-5.

A change in density will cause the number of particles per processor to fluctuate. From our formula, we see that the total time will vary as $N^2$ with a density fluctuation. The effect of changing the number of processors can also be measured here. Suppose that the number of PNs is doubled while the cell sizes remain fixed. The effect of this is to chop the region assigned to the original PN into two equal rectangles with half as many cells as before. The number of interactions on each processor is then cut in half. Since the interaction calculation tends to dominate the overall iteration time, the effect of this is to cut the iteration time in half.

In summary, the model tells us that we should make as many cells as possible on each processor. It also tells us that doubling the number of processors will cut the iteration time in half. These observations will be analyzed further with our actual timing results.

## 5.2 Timing results

To time the MD algorithm, a variety of test runs were performed using a large lattice of particles with a density of $\rho = N/\sigma^d = 1.0$ ($d$ is the dimension). These were run on a wide variety of CM-5 partitions ranging from 32 to 1024 PNs. All calculations were performed using *double precision* unless otherwise indicated.

### 5.2.1 2D timings

Using a fixed number of cells per processor, the timings in Table 1 were obtained. The update time is the time to complete one integration step.

Table 1 : Update times (sec). 36 cells per processor (fixed)

| Particles | Processors | | | | | |
|---|---|---|---|---|---|---|
| | 32 | 64 | 128 | 256 | 512 | 1024 |
| 16384 | 0.21 | 0.07 | * | * | * | * |
| 65536 | 2.70 | 0.70 | 0.20 | 0.07 | * | * |
| 262144 | 43.50 | 10.96 | 2.57 | 0.68 | 0.19 | 0.07 |
| 1048576 | 691.27 | 173.65 | 44.16 | 10.98 | 1.54 | 0.82 |
| 4194304 | - | - | 691.81 | 174.36 | 44.96 | 12.16 |
| * Cell size would be less than cut-off distance. | | | | | | |

With a fixed number of cells, the update time scales quadratically with the number of particles. For large numbers of particles, the performance degrades remarkably. This effect is the motivation for using a varied number cells. Table 2 shows the timing when a fixed cell size is used, but the number of cells is varied according to the overall size of the system.

Table 2 : 2D Update Times (sec). Cell size = $2\sigma \times 2\sigma$

|           | Processors | | | | | |
|----------:|------|------|------|------|------|------|
| Particles | 32   | 64   | 128  | 256  | 512  | 1024 |
| 16384     | 0.11 | 0.05 | 0.03 | 0.02 | 0.01 | 0.006|
| 65536     | 0.33 | 0.18 | 0.10 | 0.05 | 0.03 | 0.02 |
| 262144    | 1.24 | 0.65 | 0.34 | 0.19 | 0.08 | 0.05 |
| 1048576   | 4.96 | 2.42 | 1.15 | 0.63 | 0.34 | 0.19 |
| 4194304   | -    | 9.44 | 4.80 | 2.27 | 1.17 | 0.69 |
| 16777216  | -    | -    | -    | 9.29 | 4.77 | 2.56 |
| 33554432  | -    | -    | -    | -    | 9.22 | 4.84 |
| 67108864  | -    | -    | -    | -    | -    | 8.83 |
| 100000000 | -    | -    | -    | -    | -    | 18.57|

Here, the varied choice of cell size makes a dramatic difference. The original time for a million particles on a 32 node machine was 691 seconds, but by increasing the number of cells, it has dropped to 4.9 seconds–a factor of about 140 times faster. For a fixed partition size, the execution times scale linearly with the number of particles. When the number of processors is doubled for a given number of particles, the execution time is cut in half as predicted by the earlier analysis.

It is often useful to measure the execution time in terms of time/particle update. Dividing the times in the above table by the number of particles, we get the times in Table 3 :

Table 3 : 2D Update times/particle ($\mu$sec). Cell size = $2\sigma \times 2\sigma$

|           | Processors | | | | | |
|----------:|------|------|------|------|------|------|
| Particles | 32   | 64   | 128  | 256  | 512  | 1024 |
| 16384     | 6.653| 3.174| 1.831| 0.976| 0.610| 0.348|
| 65536     | 4.990| 2.731| 1.587| 0.717| 0.457| 0.244|
| 262144    | 4.734| 2.495| 1.312| 0.717| 0.313| 0.202|
| 1048576   | 4.728| 2.303| 1.010| 0.605| 0.328| 0.182|
| 4194304   | -    | 2.250| 1.143| 0.541| 0.280| 0.164|
| 16777216  | -    | -    | -    | 0.553| 0.284| 0.152|
| 33554432  | -    | -    | -    | -    | 0.275| 0.144|
| 67108864  | -    | -    | -    | -    | -    | 0.131|
| 100000000 | -    | -    | -    | -    | -    | 0.185|

The update times per particle remain relatively constant although there seems to be a slight improvement as the number of particles in increased.

Timings were also made to see how much time is spent in computation, communications, and redistribution. These are shown in Table 4.

Table 4 : 2D Timing breakdown for 1024 PNs

| Particles | Computation | Communication | Redistribution |
|----------:|-------------|---------------|----------------|
| 262144    | 68.4%       | 27.0%         | 4.6%           |
| 1048576   | 75.2%       | 20.0%         | 4.8%           |
| 4194304   | 82.5%       | 12.5%         | 5.0%           |
| 16777216  | 85.2%       | 9.9%          | 4.9%           |
| 67108864  | 87.0%       | 7.7%          | 5.3%           |

We see that the algorithm is dominated by computation of forces. The redistribution is a minimal amount of the overall iteration time. It also seems that the percentage of time in computation tends to increase as larger numbers of particles are modeled.

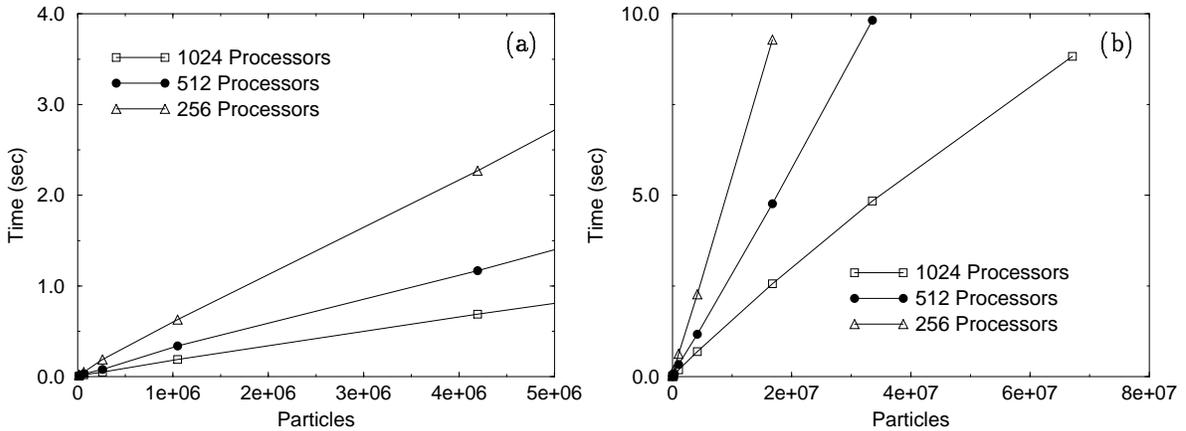

*Figure 3 : 2D times vs. number of particles.*

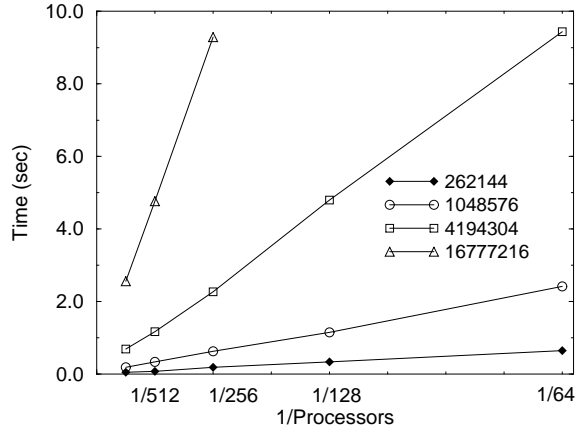

*Figure 4 : 2D scaling by number of processors.*

Fig. 3a shows the scaling by the number of particles in 2-d for several different numbers of processing nodes. Fig. 3b shows the same data over a larger range of particles. Fig. 4 shows how the iteration time scales for different numbers of processors for a given number of particles.

### 5.2.2 Verifying the scaling model

The scaling model given earlier should give reasonable results regarding the behavior of the algorithm if the constants can be determined. The constants can be determined using fixed hardware performance specifications and empirical observations. For a CM-5 without VUs, reasonable values of the constants are given below :

$$\begin{aligned}
\alpha &= 9 \times 10^{-6} \text{ sec/interaction} \\
\delta &= 7 \text{ Mbytes/sec} \\
\gamma &= 100 \mu\text{sec} \\
\beta &= 60 \text{ bytes}
\end{aligned}$$

$\alpha$ is the most difficult constant to determine. Our determination involved measuring the calculation rate in Flops for our code, and determining the average computation speed of each processing node. The number of floating point operations per force calculation is then determined and the time per interaction can then be approximated from this data. We obtained speeds of approximately 1.8 GFlops on the full CM-5 with approximately 16 floating point operations per interaction. These quantities were used to determine the value of $\alpha$ given above. The transfer rate of 7 Mbytes/sec was also a measured quantity.

Using these constants, we compare the scaling model with actual results for various runs on the 1024 PN partition. For the model, one sets $N$ to the number of particles/processor. In this case, the number of particles is divided by the number of nodes. The results are as follows:

Table 5 : 2D update times for 1024 PNs

| Particles | Actual | Scaling Model |
|----------:|-------:|--------------:|
| 16384     | 0.0057 | 0.0040        |
| 65536     | 0.016  | 0.013         |
| 262144    | 0.053  | 0.047         |
| 1048576   | 0.19   | 0.17          |
| 4194304   | 0.69   | 0.67          |
| 16777216  | 2.56   | 2.63          |
| 67108864  | 8.83   | 10.42         |

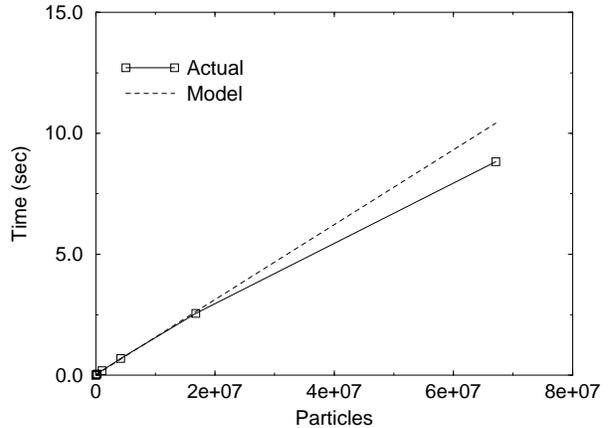

*Figure 5. Scaling model vs. observed times*

The model and actual results exhibit the same behavior and the model provides a remarkably close approximation to the execution time over a wide range of particles spanning over three orders of magnitude.

As mentioned earlier, the scaling model suggested that the best iteration time would be obtained by making the number of cells as large as possible (without violating the cut-off distance $r_{max}$). To verify this, consider a run of 1048576 particles on a 64 PN partition. In table 1 with 36 cells ($c = 6$) per PN, the iteration time is 173.65 seconds. If we use our model, this corresponds to approximately 16384 particles/PN and a predicted time of 302 seconds. In table 2 the same run is performed with a fixed cell size of $2\sigma \times 2\sigma$. This run was done with 4096 ($c = 64$) cells per PN. The measured iteration time is now 2.30 seconds and the model predicts a time of 2.63 seconds. While our scaling model does not provide a perfect fit, the same dramatic decrease in iteration time is seen by both the model and the actual code. In this case a speed improvement by a factor of 75 is obtained!

One interesting aspect of the model is the role of communications rates. Consider the 1048576 particle run on the 1024 node CM-5 given above. If we drop the communications rate to 500 Kbytes/sec, the model now predicts an interaction time of 0.216 sec. This is an increase of only 25% even though the communications rate was cut by a factor of 14. This may imply that our algorithm would effectively run on a cluster of workstations where the communication rates would be much slower than on the CM-5.

### 5.2.3  3D timings

Since the best timings in 2D occur when the cell size is held fixed, the 3D code should also have the its best times with a similar setup. Timings for the 3D algorithm with a fixed cell size are given below :

Table 6: 3D Update Times (sec). Cell size = $2\sigma \times 2\sigma \times 2\sigma$

| Particles | Processors | | | | | |
|---|---|---|---|---|---|---|
| | 32 | 64 | 128 | 256 | 512 | 1024 |
| 16384 | 0.485 | 0.259 | 0.179 | 0.098 | - | - |
| 65536 | 3.355 | 1.061 | 0.517 | 0.300 | 0.182 | 0.088 |
| 262144 | 7.683 | 3.603 | 1.992 | 0.973 | 0.512 | 0.286 |
| 1048576 | 29.566 | 15.841 | 7.007 | 3.820 | 1.983 | 1.002 |
| 2097152 | - | - | 17.864 | 8.807 | 3.812 | 1.981 |
| 4194304 | - | - | - | 20.559 | 8.711 | 3.357 |
| 16777216 | - | - | - | - | 33.630 | 19.397 |
| 67108864 | - | - | - | - | - | 87.413 |

The timings show the same behavior as before. Doubling the number of processors cuts the iteration time in half and the times scale linearly with the number of particles on each partition.

The timing breakdown of 3D is given in Table 7.

Table 7 : 3D Timing Breakdown for 1024 PNs

| Particles | Computation | Communication | Redistribution |
|---|---|---|---|
| 262144 | 70.0% | 28.5% | 1.5% |
| 1048576 | 80.0% | 19.1% | 0.9% |
| 4194304 | 82.4% | 16.6% | 1.0% |
| 16777217 | 64.7% | 34.5% | 0.8% |
| 67108864 | 58.7% | 40.6% | 0.7% |

The 3D algorithm is also dominated by computation. There is significant drop in the percentage after a certain point however. This may be due to a problem (soon to be fixed) with the time-sharing on the CM-5 (see the next section). In 3D, the amount of message passing is significantly more than in 2D and each iteration can require many thousands of message-passes. This increased message passing makes the algorithm more likely to encounter this timing problem.

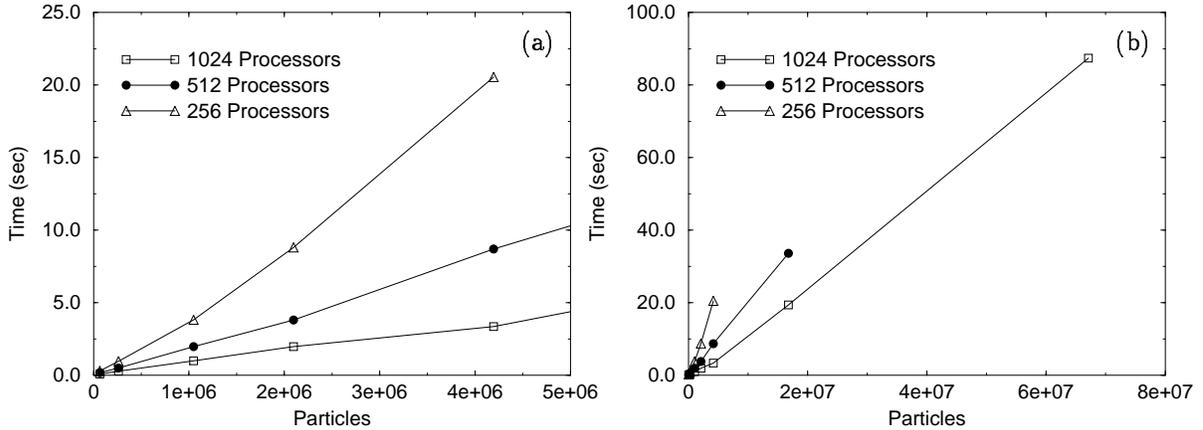

*Figure 6. Number of particles vs time in 3D*

Fig. 6a shows the scaling by the number of particles in 3D for several different numbers of processing nodes. Fig. 6b shows the same data over a larger range of particles.

A 3D scaling model can be developed in the same was as the 2D model. The corresponding 3D equation is given by

$$T = \left[\frac{27N^2}{2c^3} - \frac{N}{2}\right]\alpha + (16c^2 - 3c + 1)\left[\frac{N\beta}{c^3\delta} + \gamma\right]. \qquad (10)$$

It is interesting to note that the amount of message-passing is now a quadratic function of $c$. This indicates a dramatic increase in the amount of message-passing involved over our 2D simulations.

## 5.3 Timing summary

We have found our algorithm to be effective in modeling millions to tens of millions of particles in both 2D and 3D. Initial timings indicate that update times of approximately 0.15 $\mu$sec/particle in 2D and 1 $\mu$sec/particle in 3D are possible on a 1024 PN CM-5. It now seems possible to model at least 100 million particles in 2D and perhaps 100 million particles in 3D. We have performed several test runs to determine the limits of our code. Using single precision, we have modeled 250 million particles in 2D with an update time of 67 seconds per time-steps. With future improvements in memory and processing power, modeling 250 million particles in full precision should be within reach.

Timings from test runs are always questionable when one tries to model a problems with strong inhomogeneities. We have performed various experiments involving several million particles. Timings from these experiments are sometimes slower than the timings from our test runs, but the overall performance is quite comparable. In most cases the code runs no more than a few times slower. This degradation in performance is mainly due to the fact that it is difficult to always fully optimize our use of the PNs for every experiment. This is especially true if the problem involves a complicated geometry. It may occur that several PNs will sit idle with no particles while others will have many particles. In any case, the code has been effective in simulating significantly more complicated systems of particles than what was done in our timing runs.

We would also like to point out a slight problem with our timings. We experienced some problems obtaining accurate timings due to a problem in timesharing on the CM-5[1]. We have tried to eliminate the effects of this from our timings as much as possible, but it may have a more pronounced contribution to our runs involving very large numbers of particles.

## 6 Concluding remarks

We have implemented a scalable multi-cell MD code on the massively parallel CM-5 and thus demonstrated that meso-scale simulations with $10^8$ are now practical. Our approach is based on the message-passing model, which we find particularly well suited for mapping the basic steps of the MD algorithm onto the the underlying multicomputer. Our algorithm scales linearly with both the number of particles and number of processors used. We also derived a simple scaling relation which predicts the actual timing quite well.

Our results were obtained without the use of the CM-5 vector units (VUs). With the addition of VUs, the peak performance (overlapped mult-add instructions) of a 1024 processor machine will increase to 128 GFlops. The use of VUs will change the code in many ways that the model may not be able to account for. For example, the force calculation could be vectorized and there may be improvements in other parts of the code. Presently, the VUs are only supported from data-parallel languages such as C* and CMF. It may be possible to run C* on each of the nodes and use a data-parallel formulation locally. This would combine message-passing and a data-parallel approach. The second possibility that we are considering is direct access of the vector chips using inline DPEAC assembler code. Work in this direction is currently in progress. Preliminary studies implementing the inner kernel of the LJ-spl potential in DPEAC indicate that the kernel can run at about 50 MFlops/PN if the vector registers are kept busy for sufficiently long time. Timing results for this work will be reported elsewhere.

We thank the Los Alamos Advanced Computing Laboratory for generous support and for making their facilities available to us. We also thank J. Erpenbeck for many useful comments on the

---

[1] The problem involves context-switching in timesharing under CMOST 7.1.4.A6. When a job is about to be swapped out, all messages in the network must be removed. Upon restarting, the messages must be reinserted into the network. Sometimes the network becomes clogged and the operating system prevents further messages from being issued during that time slice. The effect is that a program will occasionally experience delays. A fix for this bug is forthcoming in a future release of CMOST.